\documentclass[aps,pra,notitlepage,showpacs,superscriptaddress,floatfix,10pt]{revtex4-1}
\usepackage{framed}
\usepackage{graphicx}
\usepackage{amsmath}
\usepackage{epsfig}
\usepackage{helvet}
\usepackage{amssymb}
\usepackage{framed}
\usepackage{bbold}
\usepackage{subfigure}
\usepackage{color}
\usepackage{listings}
\allowdisplaybreaks[3]

\begin{document}
\title{TensorNetwork: A Library for Physics and Machine Learning}

\author{Chase Roberts}
\affiliation{Alphabet (Google) X, Mountain View, CA 94043, USA}

\author{Ashley Milsted}
\affiliation{Perimeter Institute for Theoretical Physics, Waterloo, ON, Canada}

\author{Martin Ganahl}
\affiliation{Perimeter Institute for Theoretical Physics, Waterloo, ON, Canada}

\author{Adam Zalcman}
\affiliation{Google, Inc., Santa Barbara, CA 93117, USA}

\author{Bruce Fontaine}
\affiliation{Google, Inc., Mountain View, CA 94043, USA}

\author{Yijian Zou}
\affiliation{Perimeter Institute for Theoretical Physics, Waterloo, ON, Canada}

\author{Jack Hidary}
\affiliation{Alphabet (Google) X, Mountain View, CA 94043, USA}

\author{Guifre Vidal}
\affiliation{Alphabet (Google) X, Mountain View, CA 94043, USA}
\affiliation{Perimeter Institute for Theoretical Physics, Waterloo, ON, Canada}

\author{Stefan Leichenauer}
\affiliation{Alphabet (Google) X, Mountain View, CA 94043, USA}

\begin{abstract}
TensorNetwork is an open source library for implementing tensor network algorithms~\cite{library}. Tensor networks are sparse data structures originally designed for simulating quantum many-body physics, but are currently also applied in a number of other research areas, including machine learning. We demonstrate the use of the API with applications both physics and machine learning, with details appearing in companion papers.
\end{abstract}

\maketitle

\section{Introduction}

Tensor networks are sparse data structures engineered for the efficient representation and manipulation of very high-dimensional data. They have largely been developed and used in condensed matter physics~\cite{Fannes, White, Vidal, Perez-Garcia, MERA, MERA2, MERAalgorithms, Shi, Tagliacozzo, Murg, PEPS1, PEPS2, PEPS3, rev1, rev2, rev3, rev4, rev5, MERA, MERA2, MERAalgorithms}, quantum chemistry~\cite{QC1, QC2, QC3, QC4}, statistical mechanics~\cite{CTMRG, TRG, TEFRG, TNR}, quantum field theory~\cite{cMPS, cMERA}, and even quantum gravity and cosmology~\cite{Swingle, dS1, dS2, dS3, MERAgeometry}.

Substantial progress has been made recently in applying tensor networks to machine learning. Stoudenmire and Schwab used a matrix product state (MPS) for classification of the MNIST dataset~\cite{ML1}. Levine et al. showed that a deep convolutional arithmetic circuit (ConvAC) is equivalent to a tree tensor network, and gave empirical support for a more general relationship between tensor networks and convolutional network architectures~\cite{ML3}. Liu et al. employed a two-dimensional hierarchical tree tensor network for image recognition on both MNIST and CIFAR-10~\cite{ML9}. These are only a few examples from the growing body of literature on tensor networks and machine learning (for more, see e.g.~\cite{ML2, ML4, ML5, ML6, ML7, ML8, ML10, ML11}). One of our primary goals in creating the TensorNetwork library is to accelerate this research.

To keep this document self-contained, we will review the basics of tensor networks in Section~\ref{sec-overview}. Then in Section~\ref{sec-api} we will describe how to perform the most common tensor network computations using the TensorNetwork API, which is built on top of Tensorflow~\cite{TensorFlow}. A pair of companion papers will describe sample applications in physics and machine learning, respectively, in greater detail.

\section{An Overview of Tensor Networks}\label{sec-overview}

\begin{figure}
	\includegraphics[width=0.6\textwidth]{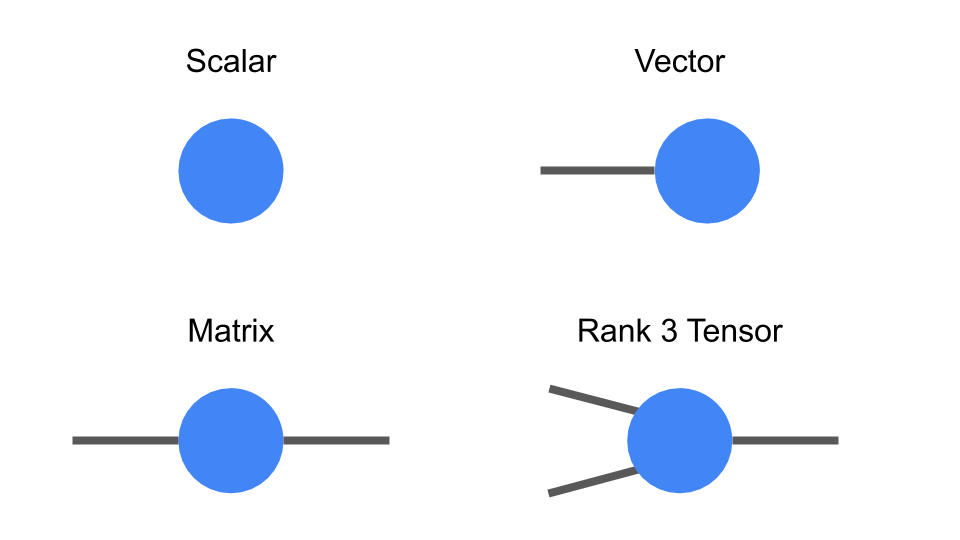}
	\caption{Examples of tensors in the graphical notation of tensor networks.}\label{fig-examples}
\end{figure}

In this section we will introduce the graphical notation for tensor networks. For our purposes a ``tensor" is synonymous with a multidimensional matrix or multidimensional array. That is, a tensor $A$ with rank $r$ consists of a set of numbers $A_{i_1,\ldots i_r}$, where the index $i_k$ has $\text{dim}_k$ possible values. The special case $r=0$ is just a number, also called a scalar. The case $r=1$ is a vector. The case $r=2$ is an ordinary matrix. Graphically, a rank $r$ tensor is represented by a node (for us a colored circle) with $r$ lines, or ``legs," coming out of it. Each of those lines represents an index. See Figure~\ref{fig-examples} for examples.

\begin{figure}
	\includegraphics[width=0.6\textwidth]{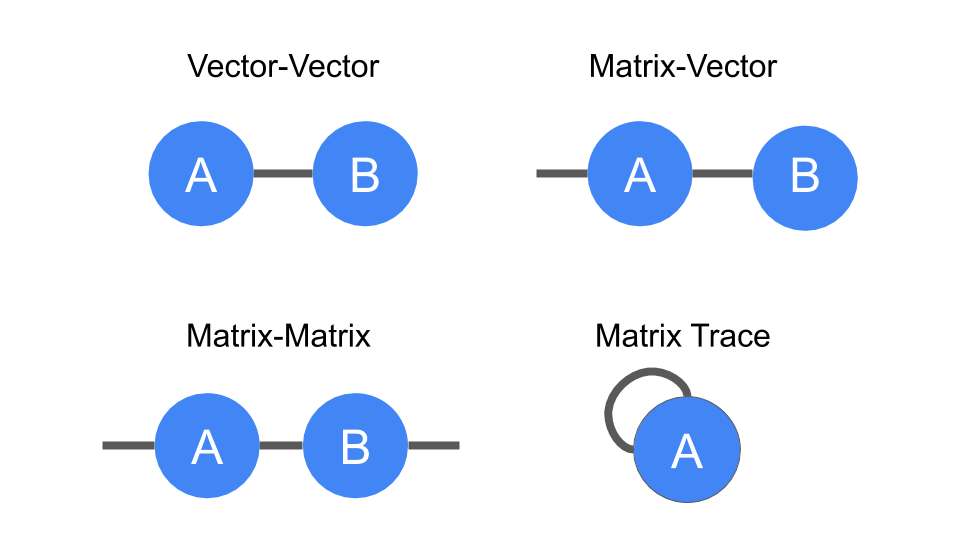}
	\caption{The graphical representations of vector-vector multiplication ($\sum_i A_i B_i$), matrix-vector multiplication ( $\sum_j A_{ij}B_j$), matrix-matrix multiplication ($\sum_j A_{ij}B_{jk}$), and matrix trace ($\sum_i A_{ii}$).}\label{fig-contractions}
\end{figure}

Now we will discuss the contraction of tensors, which is best illustrated by example. See Figure~\ref{fig-contractions} for a summary. If we are given two vectors $A$ and $B$ with the same dimension, then we can form the inner product $A\cdot B = \sum_i A_i B_i$. Notationally, we represent this by connecting the leg of $A$ with the leg of $B$. When we do this, the resulting object has no free legs. This reflects the fact that there are no remaining indices in the product $A\cdot B$. In other words, $A\cdot B$ is a scalar.

If $A$ is a matrix and $B$ is a vector where the second dimension of $A$ is equal to the dimension of $B$, then we can do the matrix-vector multiplication $\sum_j A_{ij}B_j$. Graphically, $A$ has two legs while $B$ has one, and $\sum_j A_{ij}B_j$ is represented by connecting the second leg of $A$ to the leg of $B$. The resulting object has a single leg remaining, corresponding to the one free index of $\sum_j A_{ij}B_j$. In other words, this object is a vector.

Similarly, if $A$ is a matrix and $B$ is a matrix where the second dimension of $A$ is equal to the first dimension of $B$, then we can do the matrix-matrix multiplication $\sum_j A_{ij}B_{jk}$.  The resulting object is a matrix with two free legs --- a matrix.

If a single tensor has multiple legs, we can also connect them to each other if the dimensions match. For example, we can connect the two legs of a square matrix $A$ to each other. This corresponds to the scalar $\sum_i A_{ii}$, which is just the trace $\text{Tr}A$.

\begin{figure}
\setcounter{subfigure}{0}
	\subfigure[~Matrix Product State (MPS)]{\includegraphics[width=0.35\textwidth]{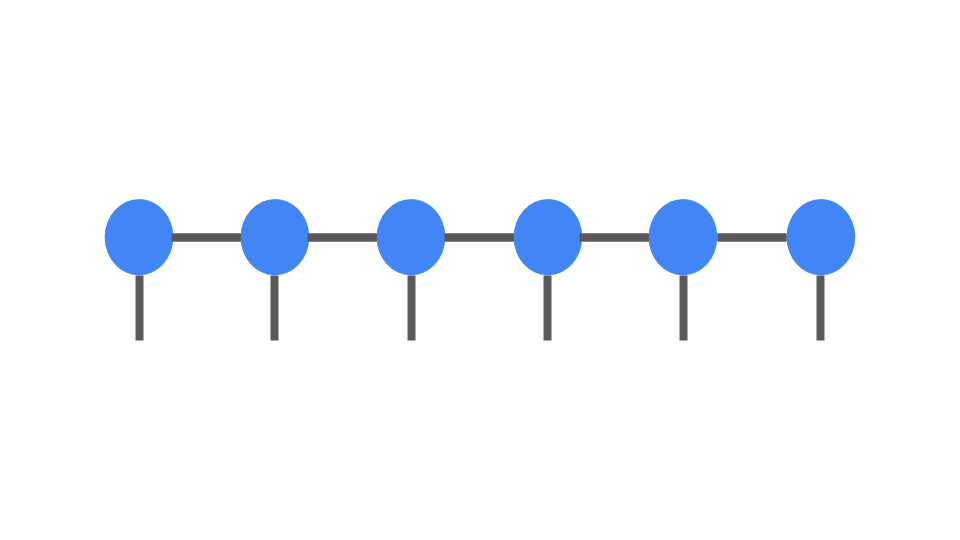}}
	\subfigure[~Tree Tensor Network (TTN)]{\includegraphics[width=0.35\textwidth]{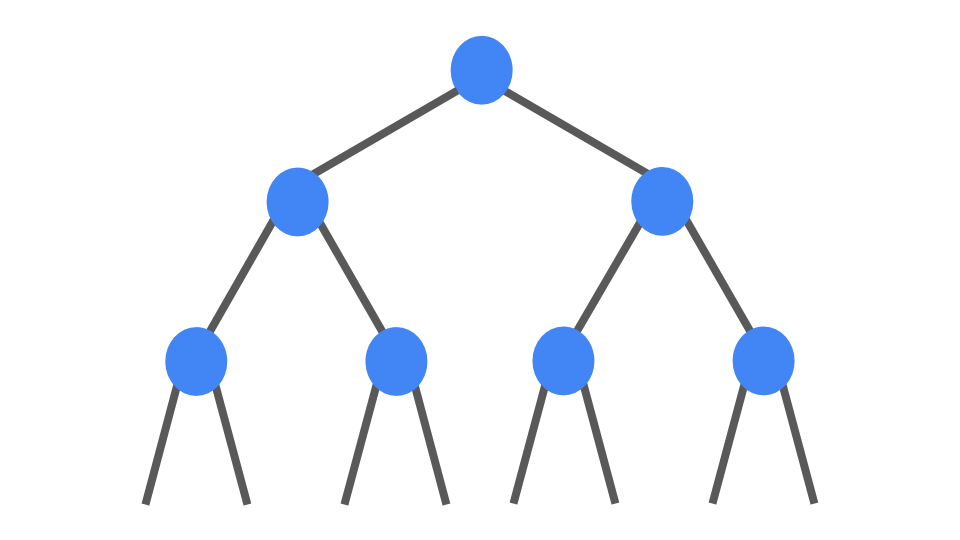}}
	\subfigure[~Multiscale Entanglement Renormalization Ansatz (MERA)]{\includegraphics[width=0.35\textwidth]{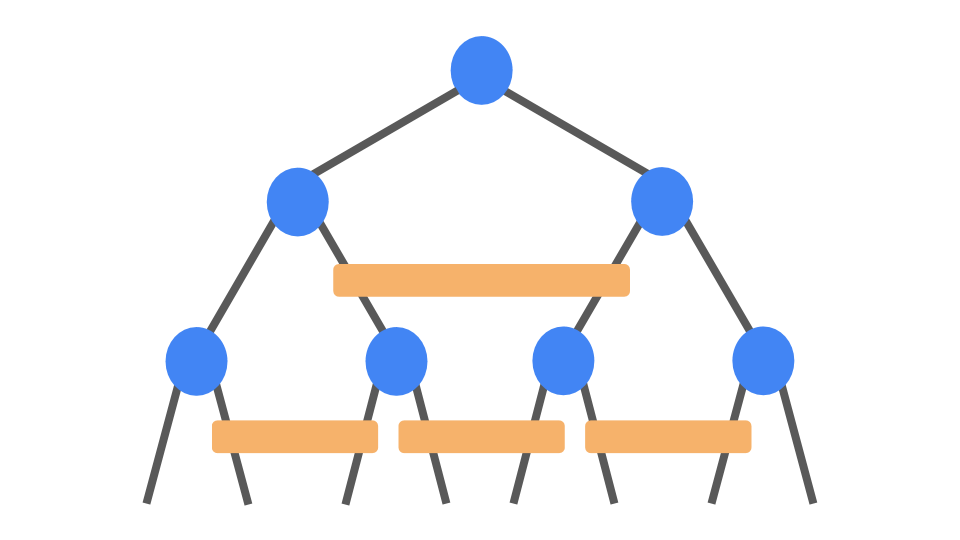}}
\caption{Three nontrivial tensor networks that are commonly used in applications. Colored shapes represent tensors and the edges connecting them represent contractions.}\label{fig-networks}
\end{figure}

Moving beyond these elementary examples, a useful tensor network is made by connecting several nodes together as part of a larger architecture. The result is a total network with many open legs that is made from the contraction of several smaller tensors. The network itself can be viewed as a very high-dimensional array, but a very sparse one. The tensor network is an efficient representation of that sparse data, and by intelligently manipulating the tensor network we can effectively deal with the high-dimensional data without a large cost. We display three examples of more advanced tensor networks in Figure~\ref{fig-networks}. 

\section{The TensorNetwork API}\label{sec-api}

In this section we describe the main ingredients of the TensorNetwork API. The functionality of the API is designed to closely mimic the manipulation of the graphical representation of a tensor network as described in Section~\ref{sec-overview}. That is, the basic objects are the nodes and edges of a graph that represents a tensor network, and the essential function of the API is to allow the user to define these objects and contract them together within networks.

\subsection{Basic Objects}

\subsubsection{TensorNetworks}

A \lstinline{TensorNetwork} is the main object of the library. It keeps track of its own set of \lstinline{Node} objects, and contains methods to add additional nodes, connect them with edges, contract them, and manipulate them in other ways.

\subsubsection{Nodes}

Nodes are one of the basic building blocks of a tensor network. They represent a tensor in the computation. Each axis will have a corresponding edge that can possibly connect different nodes (or even the same node) together. The number of edges represents the rank of the underlying tensor. For example, a node without any edges is a scalar, a node with one edge is a vector, etc. Nodes are created within a \lstinline{TensorNetwork} by passing a tensor to the \lstinline{add_node} method.

\begin{lstlisting}
import tensornetwork
import tensorflow as tf
net = tensornetwork.TensorNetwork()
a = net.add_node(tf.eye(2)) # Numpy arrays can also be passed.
print(a.get_tensor())  # This is how you access the underlying tensor.
\end{lstlisting}

\subsubsection{Edges}

\begin{figure}
\setcounter{subfigure}{0}
	\subfigure[~Standard Edge]{\includegraphics[width=0.2\textwidth]{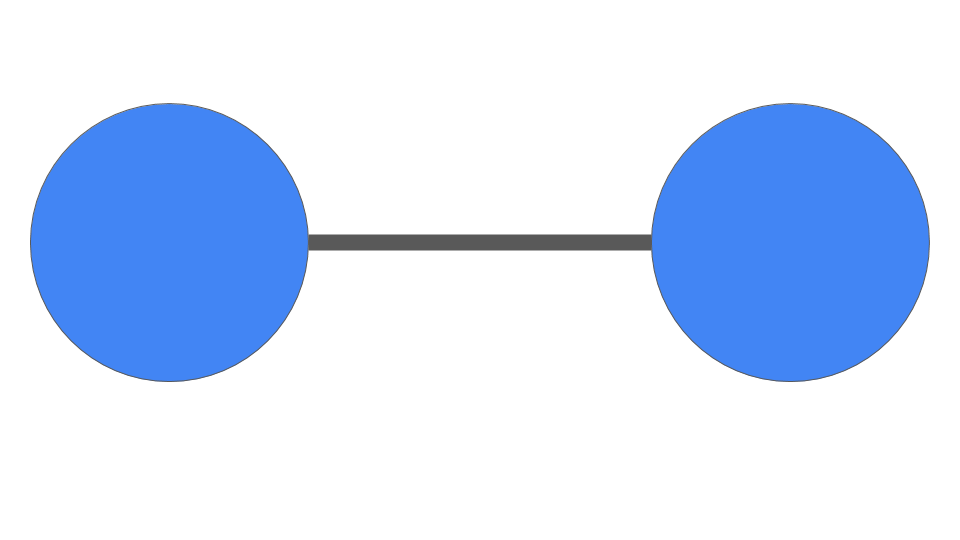}}
	\subfigure[~Trace Edge]{\includegraphics[width=0.2\textwidth]{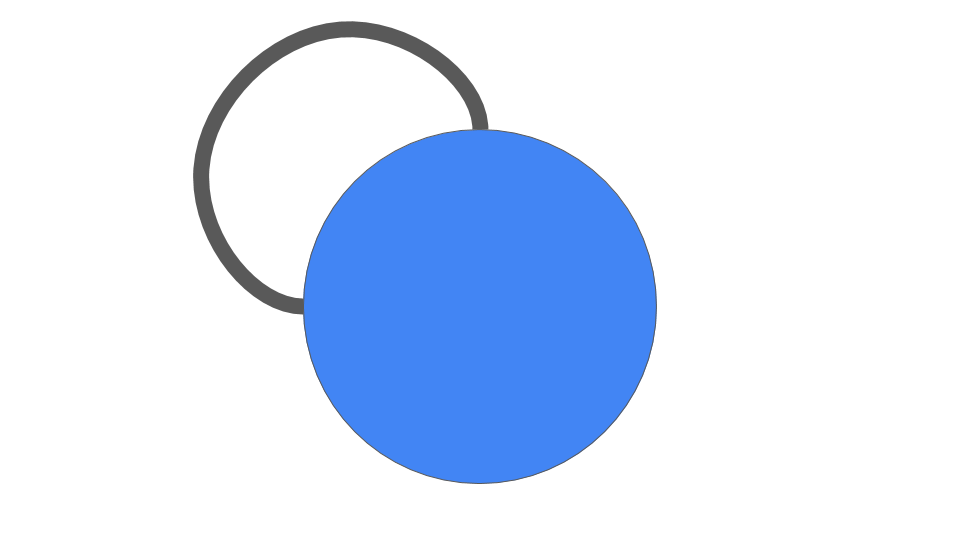}}
	\subfigure[~Dangling Edge]{\includegraphics[width=0.2\textwidth]{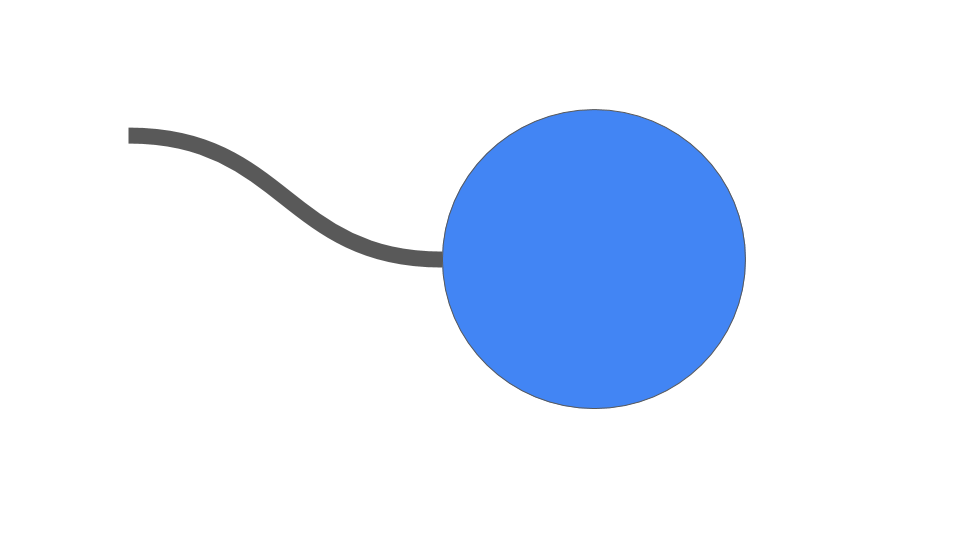}}
\caption{The three types of edges that appear in a tensor network.}
\label{fig-edges}
\end{figure}

Edges describe different contractions of the underlying tensors in the tensor network. Each edge points to the axes of the tensors to be contracted. There are three basic kinds of edges in a tensor network (see Figure~\ref{fig-edges}):

\paragraph{Standard Edges}

Standard edges are like any other edge you would find in an undirected graph. They connect two different nodes and define a dot product among the given vector spaces. In numpy terms, this edge defines a \lstinline{tensordot} operation over the given axes.

\paragraph{Trace Edges}

Trace edges connect a node to itself. To contract this type of edge, you take a trace of the matrix created by the two given axes.

\paragraph{Dangling Edges}

Dangling edges are edges that only have one side point to a node, whereas the other side is left ``dangling." These edges represent output axes or intermediate axes that have yet to be connected to other dangling edges. These edges are automatically created when adding a node to the network.

\subsection{Basic Operations}

\subsubsection{Connecting Dangling Edges}

\begin{figure}
	\includegraphics[width=0.4\textwidth]{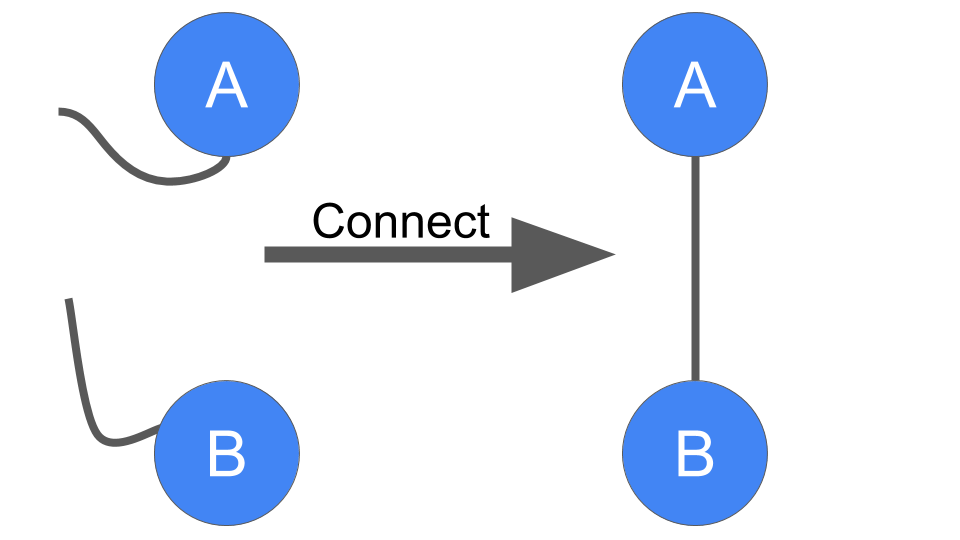}
	\caption{Connecting two dangling edges creates a new \lstinline{Edge} object between the two \lstinline{Node}s.}
	\label{fig-connect}
\end{figure}

Prior to contracting an edge, it must be connected in the network (see Figure~\ref{fig-connect}).  When constructing a tensor network, edges which begin as dangling can be connected using the \lstinline{connect} method. This method will create a new \lstinline{Edge} object which points to the two nodes that are being connected, replacing the two previous dangling edges.

We can illustrate with an example from quantum computing. Here, we have a single qubit quantum circuit where we apply a Hadamard operation several times. We connect the dangling edge of the qubit to the Hadamard operation and return the resulting ``output" edge:
\begin{lstlisting}
def apply_hadamard(net, edge):
  hadamard_op = np.array([[1.0, 1.0], [1.0, -1.0]]) / np.sqrt(2.0)
  hadamard_node = net.add_node(hadamard_op)
  # Connect the "qubit edge" to the operator "input edge" 
  net.connect(edge, hadamard_node[1])
  return hadamard_node[0]  # This is the "output edge".

# Build the quantum circuit.
net = tensornetwork.TensorNetwork()
qubit = net.add_node(np.array([1.0, 0.0])) # A "zero state" qubit.
qubit_edge = qubit.get_edge(0) # qubit[0] is equivalent.
for i in range(5):
  qubit_edge = apply_hadamard(net, qubit_edge)
\end{lstlisting}

\subsubsection{Edge Flattening}

\begin{figure}
\setcounter{subfigure}{0}
	\subfigure[~Contracting without flattening.]{\includegraphics[width=0.35\textwidth]{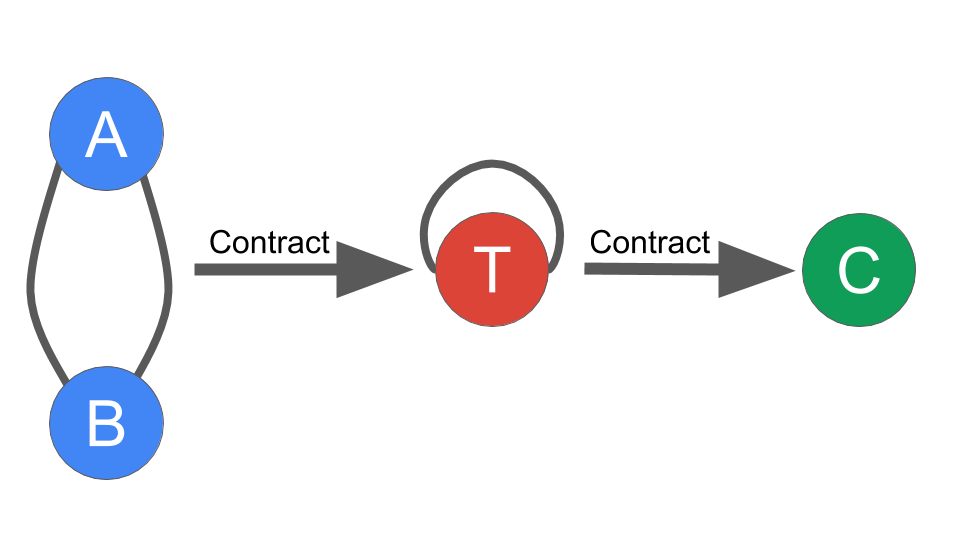}}
	\hspace{0.5in}
	\subfigure[~Flattening before contracting.]{\includegraphics[width=0.35\textwidth]{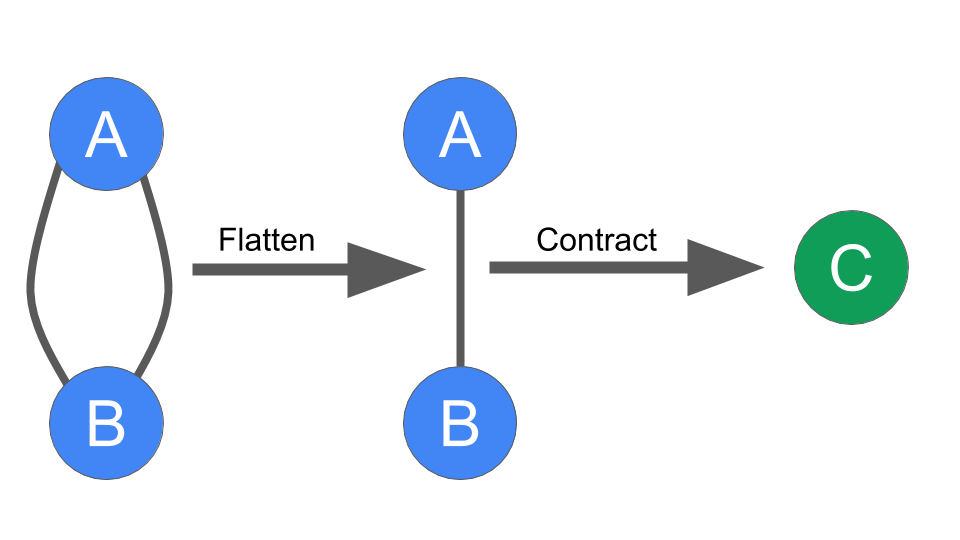}}
\caption{Contracting without flattening can produce intermediate trace edges, which results in computational overhead. Flattening before contracting is more efficient}
\label{fig-flatten-contract}
\end{figure}

It is very common for two nodes to have multiple edges connecting them. If only one of the edges is contracted at a time, then all of the remaining edges become trace edges. This is usually very bad for computation, as the new node will allocate significantly more memory than required. Also, since trace edges only sum the diagonal of the underlying matrix, all of the other values calculated during the first contraction are useless. Flattening is an effectively free operation, so it should be done every time. See Figure~\ref{fig-flatten-contract} for an illustration.

Flattening edges within the API is very simple:

\begin{lstlisting}
a = net.add_node(np.eye(2))
b = net.add_node(np.eye(2))
edge1 = net.connect(a[0], b[0])
edge2 = net.connect(a[1], b[1])
flattened_edge = net.flatten_edges([edge1, edge2])
\end{lstlisting}
All edges connecting a pair of nodes can be flattened by calling \lstinline{flatten_edges_between}:
\begin{lstlisting}
flattened_edge = net.flatten_edges_between(a, b)
\end{lstlisting}
To allow for easy pre-optimization, you can flatten all of the edges in the network with \lstinline{flatten_all_edges}. This will return all non-dangling edges in the network:
\begin{lstlisting}
contractible_edges = net.flatten_all_edges()
\end{lstlisting}

\subsubsection{Edge Contraction}

Contracting an edge is just a simple call. The tensor network API is smart enough to figure out what type of edge was passed and will do the correct computation accordingly. Here is an example which calculates the dot product of two \lstinline{ones} vectors.

\begin{lstlisting}
net = tensornetwork.TensorNetwork()
a = net.add_node(tf.ones(2))
b = net.add_node(tf.ones(2))
edge = net.connect(a[0], b[0])
c = net.contract(edge)
print(c.get_tensor().numpy()) # Should print 2.0
\end{lstlisting}
You can also automatically flatten and contract all edges connecting two nodes by calling \lstinline{contract_between}:
\begin{lstlisting}
c = net.contract_between(a, b)
\end{lstlisting}

\subsubsection{Node Outer Product}

\begin{figure}
	\includegraphics[width=0.4\textwidth]{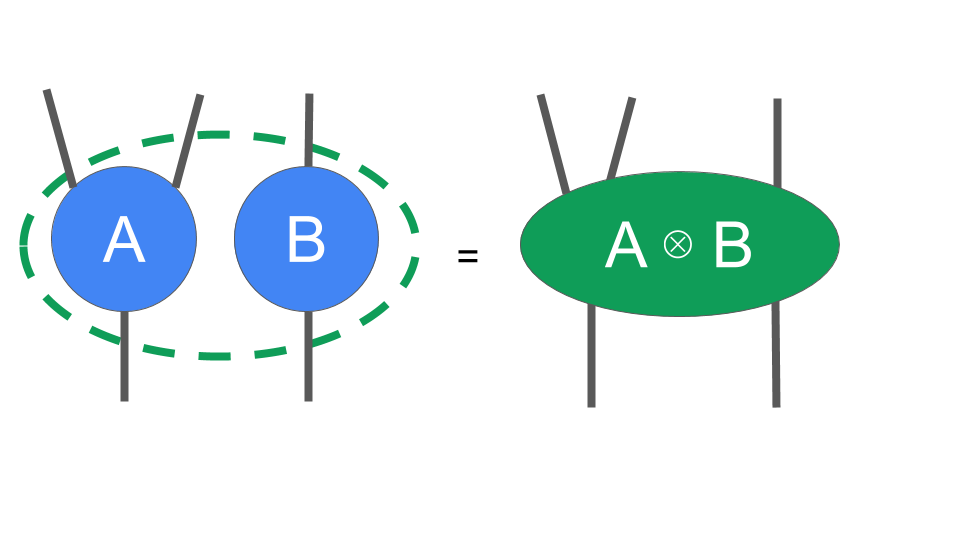}
	\caption{Two separate nodes can become one with the outer product operation}
	\label{fig-outer}
\end{figure}

The value of the outer product (also known as the tensor product or Kronecker product) of two tensors on a given assignment of indices is the element-wise product of the values of each constituent tensors with that same assignment. Explicitly written out in index notation, the outer product has the form
\[
(A\otimes B)_{i_1\cdots i_rj_1\cdots j_s} = A_{i_1\cdots i_r}B_{j_1\cdots j_s}
\]
Graphically, the outer product is simply represented by two nodes being placed next to each other and treated as a single unit (see Figure~\ref{fig-outer}). For example, if we calculated the outer product of two nodes with tensors of shape $(2, 3)$ and $(4, 5, 6)$, then the new node's tensor's shape would be $(2, 3, 4, 5, 6)$.

Doing this within the API is straightforward:
\begin{lstlisting}
c = net.outer_product(a, b)
\end{lstlisting}
This creates a new \lstinline{Node} that replaces the old \lstinline{Node}s within the network. This may seem like a strange operation to perform, since tensor networks are useful precisely because the nodes are not combined into a single higher-dimensional object. However, it is often useful to combine two nodes this way only to split them apart again with a different decomposition. That is the subject of the next section.

\subsubsection{Node Splitting}

A common step in tensor network algorithms is to do the singular value decomposition of a tensor. For example, this is used in the DMRG algorithm for calculating low-energy states of a quantum system, as well as in the use of MPS for machine learning.

For a matrix $A$, the SVD corresponds to the decomposition $A_{ij} = \sum_{k} U_{ik} S_k V_{kj}^\dagger$, where $U$ and $V$ are unitary matrices and $S$ is a vector (or diagonal matrix) with non-negative entries, the singular values. The number of nonzero singular values is at most equal to the smaller of the two dimensions of $A$.

For a more general tensor $A$ of rank $r >2$, we can generalize the SVD by splitting the legs of $A$ into two groups that form the ``effective legs" of a matrix. Each choice of groupings gives a different notion of SVD. The effective dimensions of each of these effective legs is equal to the product of the dimensions of the original legs which make it up. One can imagine reshaping the general tensor $A$ into a matrix by the procedure and then applying the ordinary matrix SVD to that matrix. Of course, we probably do not want to actually do that reshaping in practice. For notational ease, we can imagine that the ``left effective leg" consists of the first $m$ legs of $A$, and the remaining $r-m$ legs of $A$ form the ``right effective leg." In that case, the generalized tensor SVD formula becomes
\[
A_{i_1 \cdots i_r} = \sum_{k} U_{i_1\cdots i_m k} S_k V_{k i_{m+1}\cdots i_r}^\dagger.
\]
Now the number of nonzero singular values is at most the minimum of the dimensions of the two effective legs.

In our API, you can do this by calling \lstinline{split_node} or \lstinline{split_node_full_svd}. Each of these functions requires three arguments: \lstinline{node}, \lstinline{left_edges}, and \lstinline{right_edges}. The argument \lstinline{node} is just the node you want to split, \lstinline{left_edges} are the edges that you want attached to the $U$ node after splitting, and \lstinline{right_edges} are the edges that will attached to the $V^\dagger$ node after splitting. It is required that \lstinline{left_edges + right_edges} be all of the edges attached to \lstinline{node} and that none of the edges be a trace edge.

\begin{figure}
\setcounter{subfigure}{0}
	\subfigure[~\lstinline{split_node}]{\includegraphics[width=0.35\textwidth]{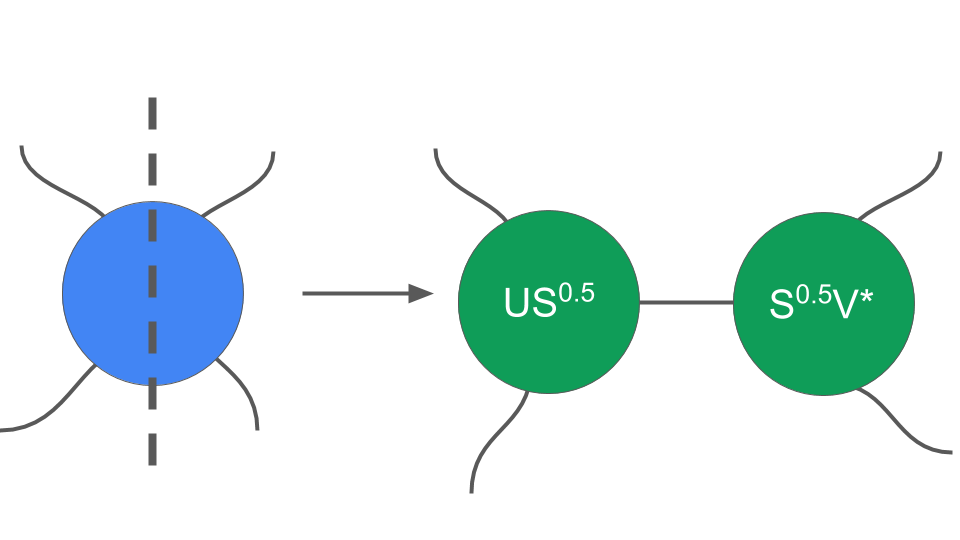}}
	\hspace{0.5in}
	\subfigure[~\lstinline{split_node_full_svd}]{\includegraphics[width=0.35\textwidth]{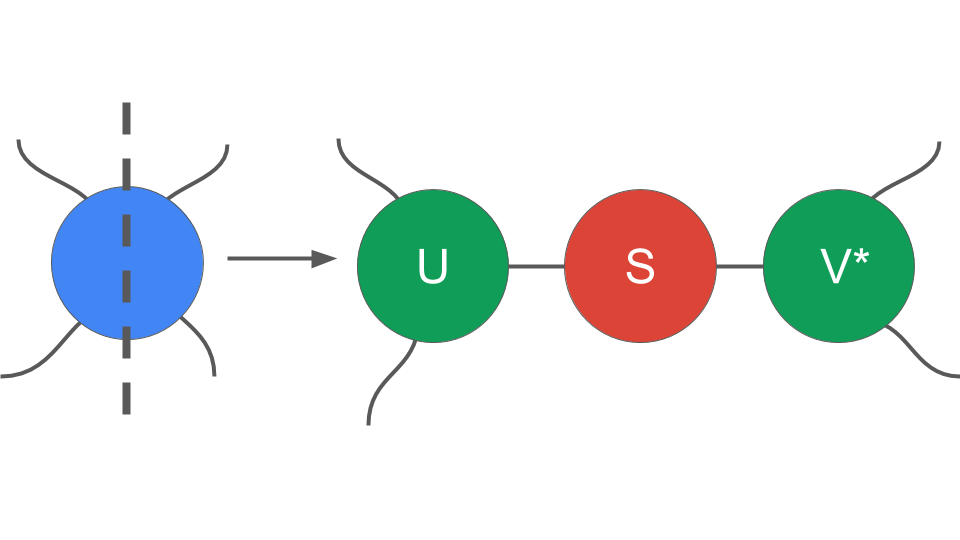}}
\caption{The two options for splitting a node treat the singular values differently. You may either absorb the square roots of the singular values into the new left and right nodes, or create a new node which stores the singular values. We include both options for convenience, even though the former option can be recovered from the latter.}
\label{fig-split}
\end{figure}

The new edge(s) created after the split will point to axes of dimension
\[
\text{new dim} = \text{min}\left(\prod \text{left edge dims}, \prod \text{right edge dims}\right)
\]
For example, if we were to split a node of shape $(2, 3, 4, 5, 6)$ with the left edges being the first three, with dimensions $(2, 3, 4)$, and the right edges being the last two, then the new left node will have shape $(2, 3, 4, 24)$ and the new right node will have shape $(24, 5, 6)$ with a new edge(s) connecting the two new axes with dimension $24$. There are two different ways that you might decide to split a single node, depending on how you wish to treat the singular values. We will explain the two options now, and illustrate them in Figure~\ref{fig-split}.

\paragraph{Split Node}

For the default \lstinline{split_node} method, the node is split into two new nodes. The values of these nodes are the unitary matrices multiplied by the square root of the singular values. It is possible to obtain a more efficient, approximate SVD by dropping the small singular values. The dropped singular values are returned in \lstinline{trun_err}, and more details about this can be found in Appendix~\ref{sec-saving}.

\begin{lstlisting}
u_s, s_vh, trun_err = net.split_node(node, left_edges, right_edges)
\end{lstlisting}

\paragraph{Split Node Full SVD}

We can get the singular values in their own separate node by using \lstinline{split_node_full_svd}. The arguments are exactly the same, only now we return three nodes instead of just two.
\begin{lstlisting}
u, s, vh, trun_err = net.split_node_full_svd(node, left_edges, right_edges)
\end{lstlisting}


\section{Use Cases}

So far we have reviewed the basics of tensor networks and how they are implemented in TensorNetwork. In a pair of companion papers~\cite{paper2, paper1}, we will dive deeper into real-world applications in both physics and machine learning. We briefly outline them here.

\subsection{Machine Learning}

Our applications of tensor networks to machine learning follows the example of Stoudenmire and Schwab~\cite{ML1}, which uses a matrix product state (Figure~\ref{fig-networks}a) to classify digits from the MNIST dataset. The method is conceptually very similar to finding the ground state energy of a physical system, except the role of ``energy" is played by a loss function. Here we will review how the image data is encoded in the tensor network, and for the details of the implementation we refer readers to~\cite{paper2}.

Tensor networks naturally operate in linear spaces of states (like the space of quantum states in physics). For image classification, this would be the space of images. One must be careful to define what the space of images is. An image consists of an ordered list of $N$ pixels, and each of those pixels has a value. For simplicity, let us assume the images are greyscale, so that a pixel's value is a single number. A naive definition of the ``space of images" would be an $N$-dimensional space, where the list of pixel values corresponding to a given image could be directly interpreted as a column vector in that space. That is \textit{not} the space of images we will use. This naive image space has the property that changing the value of a single pixel gives an almost-identical image vector. We will instead use a more powerful notion of image space that is $2^N$-dimensional, where changing the value of a single pixel can lead to a totally independent vector. 

To construct the image space, we first encode each pixel of our image into a two-dimensional pixel-space. If the image is black-and-white, so that each pixel has only two possible values, then we can use the familiar one-hot encoding into the pixel-space. That means one value of the pixel is mapped to $(1,0)^T$ and the other to $(0,1)^T$. In the more general greyscale case, we can define a \textit{local feature map} which takes each pixel value $p$ and maps it to a vector:
\[
\Phi(p) = \begin{pmatrix} \cos \pi p/2 \\ \sin{\pi p/2} \end{pmatrix}
\]
This is not the only possible feature map, but it is a particularly simple choice. Restricting to $p \in \{0,1\}$ recovers one-hot encoding for a black-and-white image. If we perform this local feature map on every pixel, then the entire image has been encoded into the tensor product of all of the two-dimensional pixel-spaces. This is our $2^N$-dimensional pixel space.

A tensor network like the MPS of Figure~\ref{fig-networks}a can be thought of as a sparse representation of a vector in the $2^N$-dimensional image space. To see that, imagine that each of the dangling legs in the network corresponds to one of the two-dimensional pixel spaces. If there are $N$ dangling legs, then the tensor network has $2^N$ components. Rather than independently specifying all $2^N$ components directly, we can specify only the components of the tensors which make up the tensor network. That is why the representation is sparse. 

If we are attempting to classify our images into one of $L$ different labels, then one strategy is to construct $L$ different MPS tensor networks which represent each of those $L$ labels. Taking the inner product of a given image vector with each of the $L$ MPS networks produces a score for that label, and the label with the highest score is assigned to the given image. The construction of these special MPS tensor networks amounts to the training of the model, the details of which can be found in the work of Stoudenmire and Schwab~\cite{ML1}, as well as in our TensorNetwork implementation~\cite{paper2}.

\subsection{Physics}

For our physics-based application we use a tree tensor network (Figure~\ref{fig-networks}b) to find the ground state of a quantum spin chain. The details of the implementation can be found in~\cite{paper1}, but here we will review the basic setup and summarize the results.

The basic idea (familiar to physicists) is very similar to the image classification setup above. The tensor network represents a vector in a state space, but in this case it is the space of possible spin configurations in a physical system, rather than the space of images. This is the setting of quantum mechanics, where the state space of any physical system is a vector space. In this context, ``training" the tensor network amounts to turning the tensors so that the resulting state has some desired property. In our case, we wish the vector to be the ground state of a particular spin system. That is, we want the vector to represent the state of lowest energy with respect to some chosen energy function.

The algorithm for training a TTN to find the ground state energy is a well-known one, and can be found in~\cite{Tagliacozzo}. A benefit of using the TensorNetwork library is that the same code can be run on multicore CPUs as well as GPUs with minimal effort. The results, reported in~\cite{paper1}, show that a 100x improvement in speed can be found by running on GPUs. This particular algorithm is well-suited for implementation in a TensorFlow-based library like TensorNetwork, since the primary computational bottleneck is tensor contractions. These are the operations for which one expects a GPU to have the greatest advantage over a CPU, and indeed we find that to be the case.


\section{Conclusion}

TensorNetwork is an open source library for TensorFlow which allows users to easily construct and manipulate tensor networks. There are multiple advantages to using TensorFlow as the basis for TensorNetwork. First, TensorFlow is already a popular and familiar tool for practitioners of machine learning. Existing software packages for tensor networks, as well as most pedagogical introductions to the subject, are phrased in the language of quantum physics. This means that, until now, the barrier to entry into the subject for the ML community has been artificially high. By using TensorFlow, we are consciously attempting to lower that barrier. This will help spur further adoption of tensor network methods for machine learning.

A second reason to use TensorFlow is that TensorFlow provides an easy way to access high-powered compute resources, including GPU and TPU clusters in the cloud. These resources are in place mainly for machine learning applications, but we aspire to leverage them for physics and chemistry as well through the TensorNetwork library.

\begin{acknowledgements}
A.~Milsted, M.~Ganahl, G.~Vidal, and Y.~Zou thank X for their hospitality. X for their hospitality. 
X is formerly known as Google[x] and is part of the Alphabet family of companies, which includes Google, Verily, Waymo, and others (www.x.company). Research at Perimeter Institute is supported by the Government of Canada through the Department of Innovation, Science and Economic Development Canada and by the Province of Ontario through the Ministry of Research, Innovation and Science.
\end{acknowledgements}

\appendix
\section{Saving Memory}\label{sec-saving}

One way to save memory during computation is to drop the lowest singular values. In  some situations, this will allow you to save an order of magnitude amount of memory while only acquiring a small amount of predictable error. You can do this easily by setting the arguments \lstinline{max_singular_values} or \lstinline{max_truncation_err}. 
The last item returned in \lstinline{split_node} and \lstinline{split_node_full_svd} is the \lstinline{truncation_error}. This is a \lstinline{tf.Tensor} of the dropped singular values. This can be useful to help keep track of the error buildup during computation.
\begin{lstlisting}
# Let's assume `node` has a tensor of shape (1000, 1000)
u, vh, trun_err = net.split_node(node, [node[0]], [node[1]], max_singular_values=5)
# u and vh are shape (1000, 5), which is 100x less memory!
print(trun_err.shape)  # Should print (995,)

# All singular values that add up to less than 0.1 will be dropped.
u, vh, trun_err = net.split_node(
    node, [node[0]], [node[1]], max_truncation_err=0.1)
print(tf.reduce_sum(trun_err))  # Should print a value <= 0.1
\end{lstlisting}

\section{Ncon}

Many tensor network practitioners use ncon~\cite{ncon}, which is a one-line way of specifying a tensor network and a sequence of contractions. We have built-in functionality for ncon for those who wish to use it:

\begin{lstlisting}
from tensornetwork_tools import ncon
a = tf.random_normal((2,2))
b = tf.random_normal((2,2))
c = ncon([a,b], [(-1,0),(0,-2)])
print(tf.norm(tf.matmul(a,b) - c)) # Should be zero
\end{lstlisting}

We also introduce the \lstinline{ncon_network} utility method, which creates a \lstinline{TensorNetwork} using the ncon API:

\begin{lstlisting}
from tensornetwork_tools import ncon_network
a = tf.random_normal((2,2))
b = tf.random_normal((2,2))
net, e_con, e_out = ncon_network([a,b], [(-1,0),(0,-2)])
for e in e_con:
    n = net.contract(e) # Contract edges in order
n.reorder_edges(e_out) # Permute final tensor as necessary
print(tf.norm(tf.matmul(a,b) - n.tensor)) # Should be zero
\end{lstlisting}

\end{document}